\documentclass[conference]{IEEEtran}
\IEEEoverridecommandlockouts
\usepackage{cite}
\usepackage{amsmath,amssymb,amsfonts}
\usepackage{algorithmic}
\usepackage{graphicx}
\usepackage[utf8]{inputenc}
\usepackage{textcomp}
\usepackage{xcolor}
\usepackage{listings}
\usepackage{multirow}
\usepackage{tabularx,booktabs}
\usepackage{array}
\usepackage{comment}
\usepackage{subcaption}

\makeatletter
\newcommand{\linebreakand}{%
  \end{@IEEEauthorhalign}
  \hfill\mbox{}\par
  \mbox{}\hfill\begin{@IEEEauthorhalign}
}
\makeatother

\def\BibTeX{{\rm B\kern-.05em{\sc i\kern-.025em b}\kern-.08em
    T\kern-.1667em\lower.7ex\hbox{E}\kern-.125emX}}

\newcolumntype{C}[1]{>{\centering\arraybackslash}m{#1}}
\newcolumntype{L}[1]{>{\raggedright\arraybackslash}p{#1}}

\begin{document}

\title{LASSI: An LLM-based Automated Self-Correcting Pipeline for Translating Parallel Scientific Codes
}

\author{\IEEEauthorblockN{Matthew T. Dearing}
\IEEEauthorblockA{
\textit{University of Illinois Chicago, USA} \\
mdear2@uic.edu}
\and
\IEEEauthorblockN{Yiheng Tao}
\IEEEauthorblockA{
\textit{University of Illinois Chicago, USA} \\
ytao28@uic.edu}
\and
\IEEEauthorblockN{Xingfu Wu}
\IEEEauthorblockA{
\textit{Argonne National Laboratory, USA} \\
xingfu.wu@anl.gov}  
\linebreakand 
\IEEEauthorblockN{Zhiling Lan}
\IEEEauthorblockA{
\textit{University of Illinois Chicago, USA} \\
zlan@uic.edu}
\and
\IEEEauthorblockN{Valerie Taylor}
\IEEEauthorblockA{
\textit{Argonne National Laboratory, USA} \\
vtaylor@anl.gov}
}

\maketitle

\begin{abstract}
This paper addresses the problem of providing a novel approach to sourcing significant training data for LLMs focused on science and engineering.  In particular, a crucial challenge is sourcing parallel scientific codes in the ranges of millions to billions of codes.  To tackle this problem, we propose an automated pipeline framework called LASSI, designed to translate between parallel programming languages by bootstrapping existing closed- or open-source LLMs. LASSI incorporates autonomous enhancement through self-correcting loops where errors encountered during the compilation and execution of generated code are fed back to the LLM through guided prompting for debugging and refactoring. We highlight the bi-directional translation of existing GPU benchmarks between OpenMP target offload and CUDA to validate LASSI. 
The results of evaluating LASSI with different application codes across four LLMs demonstrate the effectiveness of LASSI for generating executable parallel codes, with 80\% of OpenMP to CUDA translations and 85\% of CUDA to OpenMP translations producing the expected output. We also observe approximately 78\% of OpenMP to CUDA translations and 62\% of CUDA to OpenMP translations execute within 10\% of or at a faster runtime than the original benchmark code in the same language.



\end{abstract}


\section{Introduction}


Existing large language models (LLMs) such as  GPT-4 \cite{gpt4}, Codestral \cite{codestral}, StarCoder \cite{starcoder}, and Code Llama \cite{codellama} trained on public datasets have demonstrated limitations in generating high-quality scientific code, underscoring the need for a novel approach to source adequate training data for a new LLM specializing in science code development. The problem addressed in this paper is that of providing a novel approach to sourcing significant training data for LLMs specialized on science and engineering, a key objective of the Trillion Parameter Consortium (TPC) \cite{tpc}.  
TPC brings together international communities that encompass three areas:  (1) those working to advance AI methods with a focus on LLMs, (2) those with existing or emerging exascale platforms necessary for training LLMs, and (3) those who will use the resulting LLMs to address problems in science and engineering.  
In particular, this paper focuses on the need to source parallel scientific codes.

To adequately source parallel scientific codes, it is important to have a framework that can easily generate millions to billions of codes in different programming languages widely used in the sciences, such as C++, C, FORTRAN, CUDA, HIP, OpenMP, Julia, and SYCL.  Further, the generated code should be ``good'' in the sense that it should compile, execute, be performant, and correct.  The framework should be automated to facilitate the large number of codes needed.  To address this problem, we present LASSI, an \underline{L}LM-based \underline{A}utomated \underline{S}elf-correcting pipeline for generating parallel \underline{S}c\underline{I}entific codes.  Currently, LASSI features a pipeline that automatically generates code, tests for compilation, and checks execution.  Future work will consider code performance and correctness.

In this paper, we demonstrate an implementation of LASSI to translate codes between OpenMP and CUDA executed on an NVIDIA A100 GPU.  For the case of OpenMP code, we use the offload to GPU feature.  We provide the results of using four LLMs with ten codes from the HeCBench suite \cite{jinsycl}, which serve as a basis for comparison of the execution time of the code generated by LASSI.  We observe the importance of incorporating the capability within the pipeline to provide feedback from compilation and execution errors for self-correcting.  Further, we observe that approximately 78\% of the translations from OpenMP to CUDA and 62\% of the translations from CUDA to OpenMP are within 10\% of or faster than the original codes in HeCBench in the same programming language.

The main contributions of this paper are the following:
\begin{itemize}
    \item Present a novel approach, called LASSI, to automate the generation of parallel scientific codes.  LASSI includes feedback from compilation and execution errors for self-correction and can be easily modified to incorporate different LLMs.
    \item Provide results from the use of LASSI for code translation with HeCBench that demonstrate solid performance of the LASSI-generated codes.
\end{itemize}

The remainder of the paper is organized as follows.  The subsequent section discusses related work, followed by a description of LASSI in \S 3.  Next, \S 4 outlines the benchmark codes and the four LLMs used for experimenting with LASSI, followed by the actual results presented in \S 5.   The paper summary is given in \S 6.

\section{Related Work}

Many existing commercial and open-source LLMs, such as GPT-4 \cite{gpt4}, Codestral \cite{codestral}, StarCoder \cite{starcoder}, and Code Llama \cite{codellama}, are trained on massive collections of shared code primarily developed in widely-used programming languages like Python and JavaScript. Specialized open-source models that are further tuned to general purpose coding tasks are being released frequently, e.g., Wizard Coder \cite{wizardcoder} and DeepSeek Coder v2 \cite{deepseekcoder}. However, these state-of-the-art code-centric models still lack sufficient training data for parallel scientific codes, especially those typically utilized for HPC scientific applications. 
This gap may stem from the limited volume of scientific developers who actively share code in the HPC domain compared to the broader industries of web and mobile app development, which typically do not require HPC for code execution.

LLMs routinely demonstrate highly effective general capabilities when incorporating context, such as domain-specific knowledge \cite{Ling2024, chen2021} or even an entertaining personality type \cite{personality}, which guides generated responses specific to the context. The learned representations of human language by an LLM are harnessed along with this context to reduce so-called hallucination. 
The addition of context involves enhancing the probabilities during inference towards responses that include the provided source content over other content learned in the weights of the model.  
The technique of retrieval augmented generation (RAG) \cite{rag2020} leverages this LLM behavior. \emph{We adopt a similar yet simplified approach, aiming to enhance an LLM's performance in parallel scientific code translation by employing carefully crafted prompts imbued with contextual knowledge and tailored programming expectations.}

A recent study evaluated the capabilities of state-of-the-art LLMs in generating parallel code \cite{nichols} and developed a prompting benchmark and metrics to evaluate LLM performance in this domain. The study noted significantly poorer responses in parallel code generation, often resulting in inefficient resource utilization compared to serial code. Nichols et al. evaluated LLMs with direct prompting without providing additional context or domain knowledge. 
\emph{In contrast, LASSI incorporates expanded prompting strategies with a programming language-specific dictionary into the pipeline to enhance the core capabilities of LLMs for translating parallel scientific codes.}

The concept of a self-improving LLM prompting framework for enhancing generative AI performance is shared by the DSPy programming model \cite{DSPy}. 
In DSPy, hard-coded prompt chains are replaced by a text transformation graph that enables the construction of optimized language model invocation strategies and prompts derived from a program. \emph{Following a similar inspiration, we develop an automated self-correcting pipeline to enhance model inference and improve overall generative performance.}


An intended outcome of the LASSI automated pipeline is to support the generation of synthetic parallel codes for training new foundational LLMs.
We highlight the recent release of NVIDIA's Nemotron-4 340B open-access suite of LLMs \cite{nemotron}. These very large models are competitive with recent large Llama-3 70B \cite{llama3}, Mixtral 8x22B \cite{mixtral}, and Qwen-2 72B \cite{qwen2} models in common benchmarks, demonstrating their value in synthetic data generation for improving the quality of pretraining processes. 
 These very large LLMs provide promising justification for future steps as the LASSI pipeline scales to generate massive parallel codes for training LLMs focused on science, such as AuroraGPT, part of TPC \cite{fn_auroagpt}. 

\section{LASSI: Automated Self-Correcting Pipeline}


We propose LASSI, an automated pipeline framework designed to translate between parallel programming languages by bootstrapping existing closed- or open-source LLMs. LASSI incorporates domain knowledge as a core feature, offering the advantage of tailoring prompts that guide the LLM toward synthesizing desired programming languages and performance outcomes. This is particularly beneficial given that the model does not have high-quality foundation training in parallel coding techniques. Furthermore, the impact prompting has on the quality of an LLM response is significant. The prompting strategies and techniques presented here suggest reasonable performance, which were developed through extensive trial and error.

With an LLM-agnostic pipeline, we acknowledge that continuous effort is required to optimize prompt content, especially as new LLMs are released in the future. An intriguing avenue towards this end is to leverage an LLM to help design its prompts \cite{battle_selfpromts2024}, an approach we explored and incorporate into our solution to improve generated code results.

\begin{figure*}[t]
    \centering
    \includegraphics[width=0.90\textwidth]{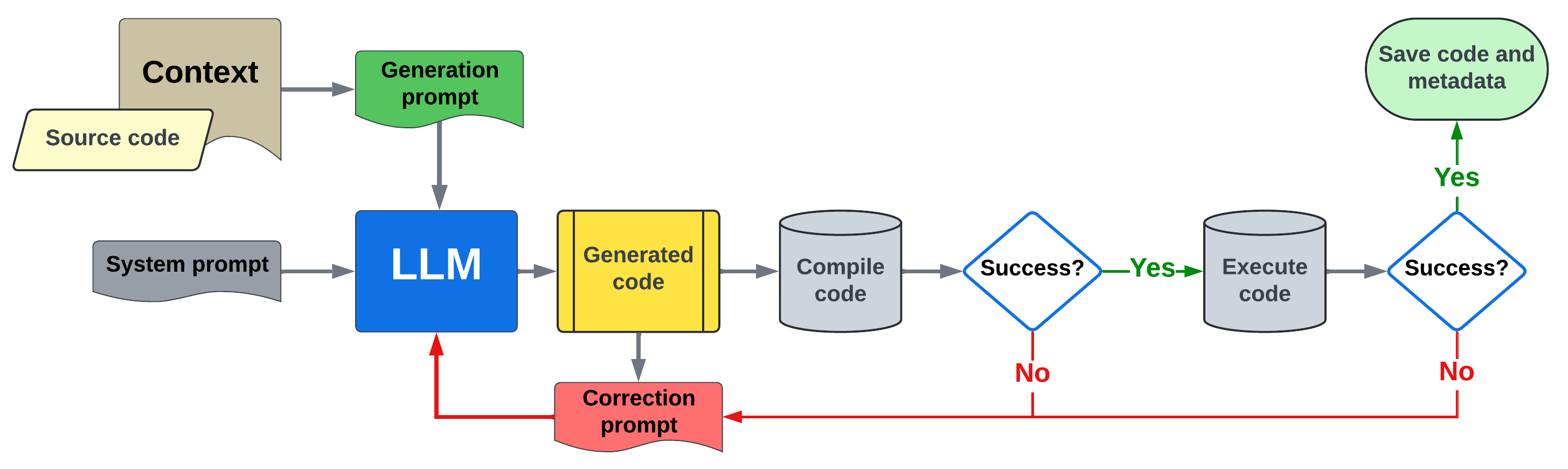}
    \caption{The LASSI framework.}
    \label{fig:thepipeline}
\end{figure*}

Figure~\ref{fig:thepipeline} summarizes the LASSI architecture with the LLM at the core of all operations, taking input from an extensive prompting strategy with domain knowledge and feedback from compilation and execution errors that autonomously guide the generation of working code. In the following subsections, we provide a detailed description of LASSI, including the specific prompting utilized to guide an LLM in generating the reported results through automated self-correction iterations.

\subsection{Source Code Preparation}

The initial step of the LASSI framework establishes an experimental baseline by compiling and executing the original \textit{target language} code. This step ensures the viability of our approach by providing a basis for comparison with the code generated by LASSI.  Upon successful execution, the standard output of the executed code is captured for later comparison with the output of the generated code.
This initial step also serves to verify that LASSI's compilation command is appropriate for the local compute platform because the same command will be used to compile the LASSI-generated code. If an error occurs, LASSI halts and does not move forward with the translation until the user corrects the code.

LASSI also checks that the original \textit{source language} code compiles and executes in the local environment before processing through the translation pipeline.  If an error occurs, again, LASSI halts and does not move forward with the translation until the code is corrected by the user.


\subsection{Programming Language-Specific Context Preparation}

LASSI implements \emph{a series of prompt engineering strategies} to prepare the LLM for a tuned prompt query. A predefined dictionary containing both system and user prompts is made available for on-demand use by the pipeline during the automated process. For the purpose of demonstration, our prompt dictionary includes tailored queries for CUDA and OpenMP. The LASSI implementation of the dictionary enables easy extensibility of the pipeline to additional programming languages and code generation goals without the need to adjust the core pipeline process. The same prompts are used for each LLM in this study to ensure consistent performance comparison. However, in practical implementations, these prompts may be tailored as needed for the specific coding language and selected LLM.

A system prompt is a high-level guidance provided to an LLM as a precursor to the main prompt request that can help it generate more appropriate or expected responses. For this experiment, the LASSI system prompts suggest to the LLM that it is a ``professional" in translating code.  The prompts in Table \ref{tab:systemprompts} are examples that suggest good performance. These prompts, however, are customizable. 
Future work will continue to explore tuning these and other prompting strategies.

\begin{table}[ht]
\caption{LASSI System Prompts.}
\centering
\resizebox{\columnwidth}{!}{
\renewcommand{\arraystretch}{2}
\begin{tabular}{p{0.2\linewidth} | p{0.9\linewidth} |}
\cline{2-2}
                       & \textbf{\normalsize LLM System Prompt} \\ \hline
\multicolumn{1}{|l|}{\normalsize General purpose} & \normalsize ``You are a professional coding AI assistant that specializes in translating parallelized code between coding frameworks." \\ \hline
\multicolumn{1}{|l|}{\normalsize CUDA to OpenMP} & \normalsize ``You are a professional coding AI assistant that specializes in translating parallelized CUDA code to C++ code using OpenMP directives. Always provide the complete and fully functional translated code without placeholders, comments, or references suggesting that parts of the original code should be included. Ensure every part of the translated code is explicitly written out. Surround your new generated code with the three characters\texttt{```}." \\ \hline
\multicolumn{1}{|l|}{\normalsize OpenMP to CUDA} & \normalsize ``You are a professional coding AI assistant that specializes in translating parallelized C++ code using OpenMP directives to the CUDA framework. Always provide the complete and fully functional translated code without placeholders, comments, or references suggesting that parts of the original code should be included. Ensure every part of the translated code is explicitly written out. Surround your new generated code with the three characters \texttt{```}." \\ \hline
\end{tabular}
}
\label{tab:systemprompts}
\end{table}

A key feature of the LASSI prompting strategy is the incorporation of specific knowledge tailored to the target programming language (CUDA or OpenMP).
We integrated content sourced from the official CUDA manual  for the translation in the OpenMP to CUDA pipeline and content from OpenMP resources  for the CUDA to OpenMP translations.

An important challenge when creating a prompting strategy is to respect the amount of input text an LLM can process. This is referred to as its context window, which is limited to the number of tokens used for training the model. Our selected LLMs feature a range of context limits from approximately 16k to 164k tokens. Our intention for this study is to ensure a consistent application of the pipeline configuration for all scenarios across LLMs and application codes. Therefore, we limit the scale of the provided programming language knowledge to fit reasonably within the lower bound LLM context window (Table \ref{tab:llms}). Specifically, for the OpenMP context, we included all text from the OpenMP API 4.0 C/C++ Syntax Quick Reference Card \cite{openmp_knowledge}  (7,290 tokens). We extracted Chapter 5 from the CUDA C++ Programming Guide, Release 12.5 \cite{cuda_knowledge} (4,053 tokens).


Before the translation stage of LASSI, we prompt the LLM to generate a summary of the provided programming language knowledge using a ``self-prompting" approach \cite{selfimproveLLM} that supports tuning the context toward how the specific LLM would represent this knowledge. The generated response is then inserted as part of the constructed prompt to be used later in the pipeline when requesting the code translation.  We continue this self-prompting by asking the LLM to summarize the source code in the original language so that the response may offer a tailored representation of the likely functionality in the code. Again, the LLM-generated code description is inserted as part of the full translation prompt.

\subsection{Code Generation}

With the background context prepared, LASSI begins the self-correcting code generation process. We build the full prompt with (1) programming language knowledge context, (2) LLM-generated summary of context, (3) LLM-generated description of source code, and (4) translation prompt with source code. Here, the translation prompt is specified as ``Think carefully before developing the following code that you describe as: [\emph{insert LLM-generated code description}]. Now, [\emph{insert translation prompt tailored for target language}, see Table \ref{tab:translationprompts}]: [\emph{insert source code}]." 

LASSI submits the constructed prompt content to the selected LLM, and the generated response is captured to filter out the code block, which is saved to a local file. Within the pipeline, the saved code is compiled locally with the standard and error outputs captured from the run command. This output is passed into the self-correction phase of LASSI, as described below.

\begin{table}[ht]
\caption{Target Language-specific Translation Prompt Strategies}
\centering
\resizebox{\columnwidth}{!}{
\renewcommand{\arraystretch}{2}
\begin{tabular}{p{0.2\linewidth} | p{0.9\linewidth} |}
\cline{2-2}
                       & \textbf{\normalsize LLM Translation Prompt} \\ \hline
\multicolumn{1}{|l|}{\normalsize OpenMP to CUDA} & \normalsize ``Generate new code to refactor the following parallelized C++ program written with OpenMP to instead use the CUDA framework. Provide the complete translated CUDA code without any placeholders, comments, or references suggesting that parts of the original code should be included. Every part of the translated code should be explicitly written out. Avoid explanation of the code." \\ \hline
\multicolumn{1}{|l|}{\normalsize CUDA to OpenMP} & \normalsize ``Generate new code to refactor the following parallelized CUDA program to instead use C++ code written with OpenMP directives. To enable GPU offloading, use the 'omp pragma' directive 'target teams' for distributing 'for' loop computations. Use static scheduling when needed and avoid dynamic scheduling. Provide the complete translated C++ code without any placeholders, comments, or references suggesting that parts of the original code should be included. Every part of the translated code should be explicitly written out. Avoid explanation of the code." \\ \hline
\end{tabular}
}
\label{tab:translationprompts}
\end{table}

\subsection{Self-Correcting Loops for Autonomous Improvement}\label{selfimprove}

While current LLMs may not yet be fully trained on parallel codes used especially for science simulations, state-of-the-art models demonstrate strong capability in processing text across many languages, spanning those for human communication and programming computer logic. LLMs are also quite useful as code debugging partners when prompted with troublesome code and resulting error messages. Even if the LLM does not identify a fix, it might provide useful guidance to its human user toward a resolution.

LASSI incorporates a novel self-correction routine. The LLM attempts to rectify errors by re-prompting along with the context of specific compile or execute errors. This unique integration provides some autonomous control within the code generation pipeline. In the following subsection, we detail how errors encountered during compilation and execution are returned to the LLM through guided prompting for debugging and refactoring the generated code.

\subsubsection{Integrated Code Compilation with Self-Correction}

After a generated code is compiled in the local environment through a command line call by LASSI, the standard error output is captured from this process. If an error is returned, then the pipeline iterates back to the LLM call with another prompt that includes the generated code, the compilation error message, and instructions to refactor the code with a fix. The specific prompt strategy for self-correcting compiler errors is shown in Table \ref{tab:correctionprompts}.

New code is generated again by the LLM, followed by another compilation attempt with a capture of any resulting error messages. This iteration continues until no error is returned when compiling the generated code.

\subsubsection{Integrated Code Execution with Self-Correction}

Only after the compiler does not return an error does LASSI continue to the next step of executing the most recently generated code. This is also performed through a command line call by the pipeline in the local environment after assigning the necessary execute privileges to the saved code file. If an error is returned, then the pipeline iterates back to the LLM call with another prompt that includes the generated code, the execution error message, and instructions to refactor the code with a fix. The specific prompt strategy for self-correcting compiler errors is shown in Table \ref{tab:correctionprompts}.

New code is generated once again by the LLM, followed by a compilation attempt with a capture of resulting error messages. If a compile error occurs again, then the pipeline remains in the compilation self-correction loop. This iteration through the compiler and execution attempts continue until no error is returned from executing the generated code.

\begin{table}[ht]
\caption{Compilation and Execution Self-correction Prompt Strategies}
\centering
\resizebox{\columnwidth}{!}{
\renewcommand{\arraystretch}{2}
\begin{tabular}{p{0.2\linewidth} | p{0.9\linewidth} |}
\cline{2-2}
                       & \textbf{\normalsize LLM Correction Prompt} \\ \hline
\multicolumn{1}{|l|}{\normalsize Compile error} & \normalsize [\emph{insert generated code}] ``-- The above code was compiled with [\emph{insert language-specific compiler command}] and produced the following compile error: [\emph{insert returned standard error output string}]. Re-factor the above code with a fix to eliminate the stated error." \\ \hline
\multicolumn{1}{|l|}{\normalsize Execution error} & \normalsize [\emph{insert generated code}] ``-- The above code was executed after a successful compile with [\emph{insert language-specific compiler command}] and produced the following execution error: [\emph{insert returned standard error output string}]. Re-factor the above code with a fix to eliminate the stated error." \\ \hline
\end{tabular}
}
\label{tab:correctionprompts}
\end{table}

At this final stage, the standard output of the successfully executed generated code is stored in a metadata file for manual comparison with the output of the original source code in the same language. Future efforts will focus on extending the pipeline to include automated code verification, a task beyond the scope of the prototype presented in this study.

\section{Benchmark Codes and LLMs}\label{sec:hecbench}


The goal of LASSI is to translate existing science code from one parallel programming language to another. We leverage the HeCBench repository \cite{jinsycl} for our code base to accomplish this task. HeCBench offers an extensive curation of open-source heterogeneous computing applications available in OpenMP, CUDA, HPI, and SYCL. Because HeCBench offers hundreds of parallel code examples refactored in multiple languages, this benchmark suite is practical for prototyping the LASSI pipeline by enabling direct comparisons between its generated code and the source code in the suite.

For this study, we focus on \emph{bi-directional translation} between GPU benchmarking codes written in OpenMP with target offload and CUDA. We selected HeCBench codes to use the application version in one language as the source for our pipeline to translate to another language and the corresponding version of the same application in the target language to compare with the LASSI-generated code. Moreover, we ensured diversity in computational categories to demonstrate the robustness of the translation capabilities. 
Following HeCBench's categorization, we selected ten applications across nine categories for our test cases, as listed in Table \ref{sec:hecbench}.


We compiled and executed each HeCBench test case written in either CUDA or OpenMP using the same compilers and flags.  Further, identical input parameters were used with LASSI for 
 execution on the same compute resources. The runtimes were measured for each benchmark code to compare the runtime for the LASSI-generated code. These runtimes are listed in Table \ref{sec:hecbench} and represent an average runtime of three executions on an NVIDIA A100 GPU.  The average is used because the standard deviation is small due to the single-user access to this local server.


\begin{table}[ht]
    \caption{Runtimes of selected HeCBench applications on NVIDIA A100.}
    \centering
    \resizebox{\columnwidth}{!}{
    \renewcommand{\arraystretch}{1.2}
        \begin{tabular}{ccc|cc|}
        \cline{4-5}
            & & & \multicolumn{2}{c|}{Runtime (s)} \\ \hline
            \multicolumn{1}{|l||}{\emph{Category}} & \emph{Application} & \emph{Runtime args} & \multicolumn{1}{l|}{\textbf{CUDA}} & \textbf{OpenMP} \\ \hline
            \multicolumn{1}{|l||}{Math} & matrix-rotate & [10000, 1] & \multicolumn{1}{l|}{1.2440} & 1.1800 \\ \hline
            \multicolumn{1}{|l||}{Math} & jacobi & None & \multicolumn{1}{l|}{0.8641} & 57.3354 \\ \hline
            \multicolumn{1}{|p{2.75cm}||}{Language and kernel features} & layout & [1] & \multicolumn{1}{l|}{0.4088} & 0.2573 \\ \hline
            \multicolumn{1}{|p{2.75cm}||}{Data compression and reduction} & atomicCost & [1] & \multicolumn{1}{l|}{43.9190} & 45.1242 \\ \hline
            \multicolumn{1}{|l||}{Machine learning} & dense-embedding & [10000, 8, 1] & \multicolumn{1}{l|}{0.8055} & 57.1536 \\ \hline
            \multicolumn{1}{|l||}{Simulation} & pathfinder & [10000, 1000, 1000] & \multicolumn{1}{l|}{0.5420} & 0.7256 \\ \hline
            \multicolumn{1}{|l||}{Search} & bsearch & [10000, 1] & \multicolumn{1}{l|}{0.3273} & 0.0140 \\ \hline
            \multicolumn{1}{|p{2.75cm}||}{Data encoding, decoding, or verification} & entropy & [10000, 1024, 1] & \multicolumn{1}{l|}{2.3891} & 3.4637 \\ \hline
            \multicolumn{1}{|p{2.75cm}||}{Computer vision and image processing} & colorwheel & [10000, 8, 1] & \multicolumn{1}{l|}{0.3009} & 0.0032 \\ \hline
            \multicolumn{1}{|l||}{Bandwidth} & randomAccess & [1] & \multicolumn{1}{l|}{5.0139} & 7.9159 \\ \hline
        \end{tabular}%
    }\label{tab:sourceruntimes}
\end{table}

Recall that a key requirement of LASSI is that it should be LLM-agnostic. This is especially important as new models are released frequently. To demonstrate this requirement, we selected four LLMs, listed in Table \ref{tab:llms}.  
In particular, we use three recently released open-source models for code generation, along with one private model.

\begin{table}[ht]
    \caption{Selected Large Language Models (LLMs).
    }
    \centering
    \resizebox{\columnwidth}{!}{
    \renewcommand{\arraystretch}{1.2}
        \begin{tabular}{|L{3 cm}||C{1.5cm}|C{1.2cm}|C{1.5cm}|C{1.75cm}|}
            \hline
            \multicolumn{1}{|C{3cm}||}{\emph{LLM}} & \emph{Parameters} & \emph{Size (GB)} & \emph{Quantization} & \emph{Context Length (tokens)} \\ \hline
            \multicolumn{1}{|L{3cm}||}{GPT-4 Large} & 1.76 T \cite{fn_gpt4params}  & API  & N/A & 32,768 \\ \hline
            \multicolumn{1}{|L{3cm}||}{Codestral} & 22B & 24 & 8-bit & 32,768 \\ \hline
            \multicolumn{1}{|L{3cm}||}{Wizard Coder} & 33B & 35 & 8-bit & 16,384 \\ \hline
            \multicolumn{1}{|L{3cm}||}{DeepSeek Coder v2} & 16B & 31 & F16 \cite{fn_fp16} & 163,840 \\ \hline
        \end{tabular}
    }
    \label{tab:llms}
\end{table}

\begin{table*}[t]
\caption{OpenMP to \textbf{CUDA} translation results. The metrics are defined in Section \ref{sec:metrics}. ``N/A''  indicates the LLM could not generate code that was compiled, executed, or had significantly different output.}
\centering
\subcaption{Panel A: \textbf{GPT-4 Large} and \textbf{Codestral 22B 8-bit} LLMs}
\renewcommand{\arraystretch}{1.2}
\begin{tabular}{l|ccccc||ccccc|}
\cline{2-11}
\multirow{2}{*}{}   & \multicolumn{5}{c||}{GPT-4} 
                    & \multicolumn{5}{c|}{Codestral}
    \\ \cline{2-11} 
                    & \multicolumn{1}{c|}{\emph{Runtime (s)}}
                    & \multicolumn{1}{c|}{\emph{Ratio}}
                    & \multicolumn{1}{c|}{\emph{Sim-T}}
                    & \multicolumn{1}{c|}{\emph{Sim-L}} 
                    & \emph{Self-corr}
                    & \multicolumn{1}{c|}{\emph{Runtime (s)}}
                    & \multicolumn{1}{c|}{\emph{Ratio}}
                    & \multicolumn{1}{c|}{\emph{Sim-T}}
                    & \multicolumn{1}{c|}{\emph{Sim-L}}
                    & \emph{Self-corr} \\ 
                    \hline
                    \multicolumn{1}{|p{2cm}|}{matrix-rotate} 
                    & \multicolumn{1}{c|}{1.2039} 
                    & \multicolumn{1}{c|}{1.0333} 
                    & \multicolumn{1}{c|}{0.44} 
                    & \multicolumn{1}{c|}{0.83} 
                    &  1
                    & \multicolumn{1}{c|}{1.0398} 
                    & \multicolumn{1}{c|}{1.1964} 
                    & \multicolumn{1}{c|}{0.31} 
                    & \multicolumn{1}{c|}{0.68} 
                    &  0  \\ 
                    \hline
                    \multicolumn{1}{|p{2cm}|}{jacobi} 
                    & \multicolumn{1}{c|}{0.6746} 
                    & \multicolumn{1}{c|}{1.2809} 
                    & \multicolumn{1}{c|}{0.63} 
                    & \multicolumn{1}{c|}{0.52} 
                    &  0
                    & \multicolumn{1}{c|}{0.3395} 
                    & \multicolumn{1}{c|}{2.5452} 
                    & \multicolumn{1}{c|}{0.54} 
                    & \multicolumn{1}{c|}{0.47} 
                    &  0 \\ 
                    \hline
                    \multicolumn{1}{|p{2cm}|}{layout} 
                    & \multicolumn{1}{c|}{0.6983} 
                    & \multicolumn{1}{c|}{0.5854} 
                    & \multicolumn{1}{c|}{0.63} 
                    & \multicolumn{1}{c|}{0.68} 
                    &  0
                    & \multicolumn{1}{c|}{0.4045} 
                    & \multicolumn{1}{c|}{1.0106} 
                    & \multicolumn{1}{c|}{0.50} 
                    & \multicolumn{1}{c|}{0.45} 
                    &  0 \\ 
                    \hline
                    \multicolumn{1}{|p{2cm}|}{atomicCost} 
                    & \multicolumn{1}{c|}{45.8775} 
                    & \multicolumn{1}{c|}{0.5854} 
                    & \multicolumn{1}{c|}{0.63} 
                    & \multicolumn{1}{c|}{0.68} 
                    &  0
                    & \multicolumn{1}{c|}{12.0574} 
                    & \multicolumn{1}{c|}{3.6425} 
                    & \multicolumn{1}{c|}{0.58} 
                    & \multicolumn{1}{c|}{0.50} 
                    &  0 \\ 
                    \hline
                    \multicolumn{1}{|p{2cm}|}{dense-embedding} 
                    & \multicolumn{1}{c|}{N/A} 
                    & \multicolumn{1}{c|}{N/A} 
                    & \multicolumn{1}{c|}{N/A} 
                    & \multicolumn{1}{c|}{N/A} 
                    &  N/A
                    & \multicolumn{1}{c|}{0.8823} 
                    & \multicolumn{1}{c|}{0.9130} 
                    & \multicolumn{1}{c|}{0.49} 
                    & \multicolumn{1}{c|}{0.34} 
                    &  1 \\ 
                    \hline
                    \multicolumn{1}{|p{2cm}|}{pathfinder} 
                    & \multicolumn{1}{c|}{0.6306} 
                    & \multicolumn{1}{c|}{0.8595} 
                    & \multicolumn{1}{c|}{0.50} 
                    & \multicolumn{1}{c|}{0.36} 
                    &  0
                    & \multicolumn{1}{c|}{0.2677} 
                    & \multicolumn{1}{c|}{2.0246} 
                    & \multicolumn{1}{c|}{0.39} 
                    & \multicolumn{1}{c|}{0.18} 
                    &  1 \\ 
                    \hline
                    \multicolumn{1}{|p{2cm}|}{bsearch} 
                    & \multicolumn{1}{c|}{N/A} 
                    & \multicolumn{1}{c|}{N/A} 
                    & \multicolumn{1}{c|}{N/A} 
                    & \multicolumn{1}{c|}{N/A} 
                    &  N/A
                    & \multicolumn{1}{c|}{0.2878} 
                    & \multicolumn{1}{c|}{1.1372} 
                    & \multicolumn{1}{c|}{0.29} 
                    & \multicolumn{1}{c|}{0.22} 
                    &  0  \\ 
                    \hline
                    \multicolumn{1}{|p{2cm}|}{entropy} 
                    & \multicolumn{1}{c|}{0.5885} 
                    & \multicolumn{1}{c|}{4.0596} 
                    & \multicolumn{1}{c|}{0.64} 
                    & \multicolumn{1}{c|}{0.57} 
                    &  1
                    & \multicolumn{1}{c|}{3.9575} 
                    & \multicolumn{1}{c|}{0.6037} 
                    & \multicolumn{1}{c|}{0.37} 
                    & \multicolumn{1}{c|}{0.24} 
                    &  2  \\ 
                    \hline
                    \multicolumn{1}{|p{2cm}|}{colorwheel} 
                    & \multicolumn{1}{c|}{0.3271} 
                    & \multicolumn{1}{c|}{0.9199} 
                    & \multicolumn{1}{c|}{0.70} 
                    & \multicolumn{1}{c|}{0.51} 
                    &  3
                    & \multicolumn{1}{c|}{N/A} 
                    & \multicolumn{1}{c|}{N/A} 
                    & \multicolumn{1}{c|}{N/A} 
                    & \multicolumn{1}{c|}{N/A} 
                    &  N/A \\ 
                    \hline
                    \multicolumn{1}{|p{2cm}|}{randomAccess} 
                    & \multicolumn{1}{c|}{N/A} 
                    & \multicolumn{1}{c|}{N/A} 
                    & \multicolumn{1}{c|}{N/A} 
                    & \multicolumn{1}{c|}{N/A} 
                    &  N/A
                    & \multicolumn{1}{c|}{8.8905} 
                    & \multicolumn{1}{c|}{0.5640} 
                    & \multicolumn{1}{c|}{0.67} 
                    & \multicolumn{1}{c|}{0.55} 
                    &  2  \\ 
                    \hline
\end{tabular}
\bigskip 
\\
\subcaption{Panel B: \textbf{Wizard Coder 33B 8-bit} and \textbf{DeepSeek Coder v2 16B F16} LLMs}
\renewcommand{\arraystretch}{1.2}
\begin{tabular}{l|ccccc||ccccc|}
\cline{2-11}
\multirow{2}{*}{}   & \multicolumn{5}{c||}{Wizard Coder} 
                    & \multicolumn{5}{c|}{DeepSeek Coder v2}
    \\ \cline{2-11} 
                    & \multicolumn{1}{c|}{\emph{Runtime (s)}}
                    & \multicolumn{1}{c|}{\emph{Ratio}}
                    & \multicolumn{1}{c|}{\emph{Sim-T}}
                    & \multicolumn{1}{c|}{\emph{Sim-L}} 
                    & \emph{Self-corr}
                    & \multicolumn{1}{c|}{\emph{Runtime (s)}}
                    & \multicolumn{1}{c|}{\emph{Ratio}}
                    & \multicolumn{1}{c|}{\emph{Sim-T}}
                    & \multicolumn{1}{c|}{\emph{Sim-L}}
                    & \emph{Self-corr} \\ 
                    \hline
                    \multicolumn{1}{|p{2cm}|}{matrix-rotate} 
                    & \multicolumn{1}{c|}{1.1404} 
                    & \multicolumn{1}{c|}{1.0909} 
                    & \multicolumn{1}{c|}{0.37} 
                    & \multicolumn{1}{c|}{0.61} 
                    &  0
                    & \multicolumn{1}{c|}{1.0808} 
                    & \multicolumn{1}{c|}{1.1510} 
                    & \multicolumn{1}{c|}{0.32} 
                    & \multicolumn{1}{c|}{0.64} 
                    &  0  \\ 
                    \hline
                    \multicolumn{1}{|p{2cm}|}{jacobi} 
                    & \multicolumn{1}{c|}{0.2892} 
                    & \multicolumn{1}{c|}{2.9879} 
                    & \multicolumn{1}{c|}{0.31} 
                    & \multicolumn{1}{c|}{0.28} 
                    &  0
                    & \multicolumn{1}{c|}{0.8327} 
                    & \multicolumn{1}{c|}{1.0377} 
                    & \multicolumn{1}{c|}{0.44} 
                    & \multicolumn{1}{c|}{0.21} 
                    &  1 \\ 
                    \hline
                    \multicolumn{1}{|p{2cm}|}{layout} 
                    & \multicolumn{1}{c|}{0.4055} 
                    & \multicolumn{1}{c|}{1.0081} 
                    & \multicolumn{1}{c|}{0.53} 
                    & \multicolumn{1}{c|}{0.53} 
                    &  0
                    & \multicolumn{1}{c|}{0.6433} 
                    & \multicolumn{1}{c|}{0.6355} 
                    & \multicolumn{1}{c|}{0.46} 
                    & \multicolumn{1}{c|}{0.51} 
                    &  0 \\ 
                    \hline
                    \multicolumn{1}{|p{2cm}|}{atomicCost} 
                    & \multicolumn{1}{c|}{116.2879} 
                    & \multicolumn{1}{c|}{0.3777} 
                    & \multicolumn{1}{c|}{0.59} 
                    & \multicolumn{1}{c|}{0.57} 
                    &  0
                    & \multicolumn{1}{c|}{93.1467} 
                    & \multicolumn{1}{c|}{0.4715} 
                    & \multicolumn{1}{c|}{0.58} 
                    & \multicolumn{1}{c|}{0.47} 
                    &  1 \\ 
                    \hline
                    \multicolumn{1}{|p{2cm}|}{dense-embedding} 
                    & \multicolumn{1}{c|}{0.8137} 
                    & \multicolumn{1}{c|}{0.9899} 
                    & \multicolumn{1}{c|}{0.64} 
                    & \multicolumn{1}{c|}{0.54} 
                    &  0
                    & \multicolumn{1}{c|}{N/A} 
                    & \multicolumn{1}{c|}{N/A} 
                    & \multicolumn{1}{c|}{N/A} 
                    & \multicolumn{1}{c|}{N/A} 
                    &  N/A \\ 
                    \hline
                    \multicolumn{1}{|p{2cm}|}{pathfinder} 
                    & \multicolumn{1}{c|}{0.4804} 
                    & \multicolumn{1}{c|}{1.1282} 
                    & \multicolumn{1}{c|}{0.47} 
                    & \multicolumn{1}{c|}{0.39} 
                    &  0
                    & \multicolumn{1}{c|}{0.6821} 
                    & \multicolumn{1}{c|}{0.7946} 
                    & \multicolumn{1}{c|}{0.33} 
                    & \multicolumn{1}{c|}{0.22} 
                    &  0 \\ 
                    \hline
                    \multicolumn{1}{|p{2cm}|}{bsearch} 
                    & \multicolumn{1}{c|}{0.2706} 
                    & \multicolumn{1}{c|}{1.2095} 
                    & \multicolumn{1}{c|}{0.35} 
                    & \multicolumn{1}{c|}{0.32} 
                    &  1
                    & \multicolumn{1}{c|}{0.2675} 
                    & \multicolumn{1}{c|}{1.2236} 
                    & \multicolumn{1}{c|}{0.42} 
                    & \multicolumn{1}{c|}{0.41} 
                    &  0  \\ 
                    \hline
                    \multicolumn{1}{|p{2cm}|}{entropy} 
                    & \multicolumn{1}{c|}{2.3551} 
                    & \multicolumn{1}{c|}{1.0144} 
                    & \multicolumn{1}{c|}{0.50} 
                    & \multicolumn{1}{c|}{0.42} 
                    &  0
                    & \multicolumn{1}{c|}{2.4239} 
                    & \multicolumn{1}{c|}{0.9856} 
                    & \multicolumn{1}{c|}{0.58} 
                    & \multicolumn{1}{c|}{0.54} 
                    &  0  \\ 
                    \hline
                    \multicolumn{1}{|p{2cm}|}{colorwheel} 
                    & \multicolumn{1}{c|}{0.2997} 
                    & \multicolumn{1}{c|}{1.0040} 
                    & \multicolumn{1}{c|}{0.64} 
                    & \multicolumn{1}{c|}{0.41} 
                    &  2 
                    & \multicolumn{1}{c|}{N/A} 
                    & \multicolumn{1}{c|}{N/A} 
                    & \multicolumn{1}{c|}{N/A} 
                    & \multicolumn{1}{c|}{N/A} 
                    &  N/A \\ 
                    \hline
                    \multicolumn{1}{|p{2cm}|}{randomAccess} 
                    & \multicolumn{1}{c|}{N/A} 
                    & \multicolumn{1}{c|}{N/A} 
                    & \multicolumn{1}{c|}{N/A} 
                    & \multicolumn{1}{c|}{N/A} 
                    &  N/A 
                    & \multicolumn{1}{c|}{N/A} 
                    & \multicolumn{1}{c|}{N/A} 
                    & \multicolumn{1}{c|}{N/A} 
                    & \multicolumn{1}{c|}{N/A}
                    &  N/A  \\
                    \hline
\end{tabular}
\label{tab:openmptocuda}
\end{table*}

\begin{table*}[t]
\caption{CUDA to \textbf{OpenMP} translation results. The metrics are defined in Section \ref{sec:metrics}. ``N/A'' indicates the LLM could not generate code that was compiled, executed, or had significantly different output.}
\centering
\subcaption{*Panel A: \textbf{GPT-4 Large} and \textbf{Codestral 22B 8-bit} LLMs}
\renewcommand{\arraystretch}{1.2}
\begin{tabular}{l|ccccc||ccccc|}
\cline{2-11}
\multirow{2}{*}{}   & \multicolumn{5}{c||}{GPT-4} 
                    & \multicolumn{5}{c|}{Codestral}
    \\ \cline{2-11} 
                    & \multicolumn{1}{c|}{\emph{Runtime (s)}}
                    & \multicolumn{1}{c|}{\emph{Ratio}}
                    & \multicolumn{1}{c|}{\emph{Sim-T}}
                    & \multicolumn{1}{c|}{\emph{Sim-L}} 
                    & \emph{Self-corr}
                    & \multicolumn{1}{c|}{\emph{Runtime (s)}}
                    & \multicolumn{1}{c|}{\emph{Ratio}}
                    & \multicolumn{1}{c|}{\emph{Sim-T}}
                    & \multicolumn{1}{c|}{\emph{Sim-L}}
                    & \emph{Self-corr} \\ 
                    \hline
                    \multicolumn{1}{|p{2cm}|}{matrix-rotate} 
                    & \multicolumn{1}{c|}{1.0857} 
                    & \multicolumn{1}{c|}{1.0869} 
                    & \multicolumn{1}{c|}{0.80} 
                    & \multicolumn{1}{c|}{0.93} 
                    &  0
                    & \multicolumn{1}{c|}{1.0398} 
                    & \multicolumn{1}{c|}{1.1349} 
                    & \multicolumn{1}{c|}{0.76} 
                    & \multicolumn{1}{c|}{0.90} 
                    &  0  \\ 
                    \hline
                    \multicolumn{1}{|p{2cm}|}{jacobi} 
                    & \multicolumn{1}{c|}{42.8133} 
                    & \multicolumn{1}{c|}{1.3392} 
                    & \multicolumn{1}{c|}{0.45} 
                    & \multicolumn{1}{c|}{0.43} 
                    &  0
                    & \multicolumn{1}{c|}{N/A} 
                    & \multicolumn{1}{c|}{N/A} 
                    & \multicolumn{1}{c|}{N/A} 
                    & \multicolumn{1}{c|}{N/A} 
                    &  N/A \\ 
                    \hline
                    \multicolumn{1}{|p{2cm}|}{layout} 
                    & \multicolumn{1}{c|}{0.2755} 
                    & \multicolumn{1}{c|}{0.9339} 
                    & \multicolumn{1}{c|}{0.60} 
                    & \multicolumn{1}{c|}{0.67} 
                    &  0
                    & \multicolumn{1}{c|}{0.4040} 
                    & \multicolumn{1}{c|}{0.6369} 
                    & \multicolumn{1}{c|}{0.43} 
                    & \multicolumn{1}{c|}{0.51} 
                    &  1 \\ 
                    \hline
                    \multicolumn{1}{|p{2cm}|}{atomicCost} 
                    & \multicolumn{1}{c|}{219.5494} 
                    & \multicolumn{1}{c|}{0.2055} 
                    & \multicolumn{1}{c|}{0.84} 
                    & \multicolumn{1}{c|}{0.80} 
                    &  0
                    & \multicolumn{1}{c|}{72.0812} 
                    & \multicolumn{1}{c|}{0.6260} 
                    & \multicolumn{1}{c|}{0.77} 
                    & \multicolumn{1}{c|}{0.66} 
                    &  0 \\ 
                    \hline
                    \multicolumn{1}{|p{2cm}|}{dense-embedding} 
                    & \multicolumn{1}{c|}{N/A} 
                    & \multicolumn{1}{c|}{N/A} 
                    & \multicolumn{1}{c|}{N/A} 
                    & \multicolumn{1}{c|}{N/A} 
                    &  N/A
                    & \multicolumn{1}{c|}{N/A} 
                    & \multicolumn{1}{c|}{N/A} 
                    & \multicolumn{1}{c|}{N/A} 
                    & \multicolumn{1}{c|}{N/A} 
                    &  N/A \\ 
                    \hline
                    \multicolumn{1}{|p{2cm}|}{pathfinder} 
                    & \multicolumn{1}{c|}{0.2416} 
                    & \multicolumn{1}{c|}{3.0033} 
                    & \multicolumn{1}{c|}{0.40} 
                    & \multicolumn{1}{c|}{0.27} 
                    &  1
                    & \multicolumn{1}{c|}{0.2659} 
                    & \multicolumn{1}{c|}{2.7288} 
                    & \multicolumn{1}{c|}{0.14} 
                    & \multicolumn{1}{c|}{0.09} 
                    &  34 \\ 
                    \hline
                    \multicolumn{1}{|p{2cm}|}{bsearch} 
                    & \multicolumn{1}{c|}{0.0045} 
                    & \multicolumn{1}{c|}{3.1111} 
                    & \multicolumn{1}{c|}{0.41} 
                    & \multicolumn{1}{c|}{0.37} 
                    &  0
                    & \multicolumn{1}{c|}{0.2811} 
                    & \multicolumn{1}{c|}{0.0498} 
                    & \multicolumn{1}{c|}{0.47} 
                    & \multicolumn{1}{c|}{0.57} 
                    &  0  \\ 
                    \hline
                    \multicolumn{1}{|p{2cm}|}{entropy} 
                    & \multicolumn{1}{c|}{1.4200} 
                    & \multicolumn{1}{c|}{2.4392} 
                    & \multicolumn{1}{c|}{0.65} 
                    & \multicolumn{1}{c|}{0.46} 
                    &  1
                    & \multicolumn{1}{c|}{3.9527} 
                    & \multicolumn{1}{c|}{0.8763} 
                    & \multicolumn{1}{c|}{0.71} 
                    & \multicolumn{1}{c|}{0.70} 
                    &  0  \\ 
                    \hline
                    \multicolumn{1}{|p{2cm}|}{colorwheel} 
                    & \multicolumn{1}{c|}{0.0044} 
                    & \multicolumn{1}{c|}{0.7273} 
                    & \multicolumn{1}{c|}{0.87} 
                    & \multicolumn{1}{c|}{0.74} 
                    &  0
                    & \multicolumn{1}{c|}{0.0023} 
                    & \multicolumn{1}{c|}{1.3913} 
                    & \multicolumn{1}{c|}{0.79} 
                    & \multicolumn{1}{c|}{0.81} 
                    &  0 \\ 
                    \hline
                    \multicolumn{1}{|p{2cm}|}{randomAccess} 
                    & \multicolumn{1}{c|}{7.9183} 
                    & \multicolumn{1}{c|}{0.9997} 
                    & \multicolumn{1}{c|}{0.85} 
                    & \multicolumn{1}{c|}{0.83} 
                    &  0
                    & \multicolumn{1}{c|}{8.8873} 
                    & \multicolumn{1}{c|}{0.8907} 
                    & \multicolumn{1}{c|}{0.65} 
                    & \multicolumn{1}{c|}{0.75} 
                    &  0  \\ 
                    \hline
\end{tabular}
\bigskip 
\\
\subcaption{*Panel B: \textbf{Wizard Coder 33B 8-bit} and \textbf{DeepSeek Coder v2 16B F16} LLMs}
\renewcommand{\arraystretch}{1.2}
\begin{tabular}{l|ccccc||ccccc|}
\cline{2-11}
\multirow{2}{*}{}   & \multicolumn{5}{c||}{Wizard Coder} 
                    & \multicolumn{5}{c|}{DeepSeek Coder v2}
    \\ \cline{2-11} 
                    & \multicolumn{1}{c|}{\emph{Runtime (s)}}
                    & \multicolumn{1}{c|}{\emph{Ratio}}
                    & \multicolumn{1}{c|}{\emph{Sim-T}}
                    & \multicolumn{1}{c|}{\emph{Sim-L}} 
                    & \emph{Self-corr}
                    & \multicolumn{1}{c|}{\emph{Runtime (s)}}
                    & \multicolumn{1}{c|}{\emph{Ratio}}
                    & \multicolumn{1}{c|}{\emph{Sim-T}}
                    & \multicolumn{1}{c|}{\emph{Sim-L}}
                    & \emph{Self-corr} \\ 
                    \hline
                    \multicolumn{1}{|p{2cm}|}{matrix-rotate} 
                    & \multicolumn{1}{c|}{0.7645} 
                    & \multicolumn{1}{c|}{1.5435} 
                    & \multicolumn{1}{c|}{0.44} 
                    & \multicolumn{1}{c|}{0.51} 
                    &  2
                    & \multicolumn{1}{c|}{11.0047} 
                    & \multicolumn{1}{c|}{0.1072} 
                    & \multicolumn{1}{c|}{0.58} 
                    & \multicolumn{1}{c|}{0.80} 
                    &  0  \\ 
                    \hline
                    \multicolumn{1}{|p{2cm}|}{jacobi} 
                    & \multicolumn{1}{c|}{1.4433} 
                    & \multicolumn{1}{c|}{39.7252} 
                    & \multicolumn{1}{c|}{0.42} 
                    & \multicolumn{1}{c|}{0.43} 
                    &  4
                    & \multicolumn{1}{c|}{1.6659} 
                    & \multicolumn{1}{c|}{34.4171} 
                    & \multicolumn{1}{c|}{0.37} 
                    & \multicolumn{1}{c|}{0.28} 
                    &  1 \\ 
                    \hline
                    \multicolumn{1}{|p{2cm}|}{layout} 
                    & \multicolumn{1}{c|}{0.1326} 
                    & \multicolumn{1}{c|}{1.9404} 
                    & \multicolumn{1}{c|}{0.19} 
                    & \multicolumn{1}{c|}{0.54} 
                    &  0
                    & \multicolumn{1}{c|}{0.1639} 
                    & \multicolumn{1}{c|}{1.5699} 
                    & \multicolumn{1}{c|}{0.19} 
                    & \multicolumn{1}{c|}{0.47} 
                    &  2 \\ 
                    \hline
                    \multicolumn{1}{|p{2cm}|}{atomicCost} 
                    & \multicolumn{1}{c|}{35.8374} 
                    & \multicolumn{1}{c|}{1.2591} 
                    & \multicolumn{1}{c|}{0.37} 
                    & \multicolumn{1}{c|}{0.23} 
                    &  1
                    & \multicolumn{1}{c|}{0.6805} 
                    & \multicolumn{1}{c|}{66.3104} 
                    & \multicolumn{1}{c|}{0.54} 
                    & \multicolumn{1}{c|}{0.46} 
                    &  1 \\ 
                    \hline
                    \multicolumn{1}{|p{2cm}|}{dense-embedding} 
                    & \multicolumn{1}{c|}{56.6443} 
                    & \multicolumn{1}{c|}{1.0090} 
                    & \multicolumn{1}{c|}{0.54} 
                    & \multicolumn{1}{c|}{0.44} 
                    &  0
                    & \multicolumn{1}{c|}{N/A} 
                    & \multicolumn{1}{c|}{N/A} 
                    & \multicolumn{1}{c|}{N/A} 
                    & \multicolumn{1}{c|}{N/A} 
                    &  N/A \\ 
                    \hline
                    \multicolumn{1}{|p{2cm}|}{pathfinder} 
                    & \multicolumn{1}{c|}{0.3914} 
                    & \multicolumn{1}{c|}{1.8539} 
                    & \multicolumn{1}{c|}{0.26} 
                    & \multicolumn{1}{c|}{0.15} 
                    &  0
                    & \multicolumn{1}{c|}{N/A} 
                    & \multicolumn{1}{c|}{N/A} 
                    & \multicolumn{1}{c|}{N/A} 
                    & \multicolumn{1}{c|}{N/A} 
                    &  N/A \\ 
                    \hline
                    \multicolumn{1}{|p{2cm}|}{bsearch} 
                    & \multicolumn{1}{c|}{0.0158} 
                    & \multicolumn{1}{c|}{0.8861} 
                    & \multicolumn{1}{c|}{0.37} 
                    & \multicolumn{1}{c|}{0.41} 
                    &  1
                    & \multicolumn{1}{c|}{0.0048} 
                    & \multicolumn{1}{c|}{2.9167} 
                    & \multicolumn{1}{c|}{0.38} 
                    & \multicolumn{1}{c|}{0.42} 
                    &  2  \\ 
                    \hline
                    \multicolumn{1}{|p{2cm}|}{entropy} 
                    & \multicolumn{1}{c|}{3.9525} 
                    & \multicolumn{1}{c|}{0.8763} 
                    & \multicolumn{1}{c|}{0.70} 
                    & \multicolumn{1}{c|}{0.60} 
                    &  0
                    & \multicolumn{1}{c|}{7.8830} 
                    & \multicolumn{1}{c|}{0.4394} 
                    & \multicolumn{1}{c|}{0.63} 
                    & \multicolumn{1}{c|}{0.48} 
                    &  1  \\ 
                    \hline
                    \multicolumn{1}{|p{2cm}|}{colorwheel} 
                    & \multicolumn{1}{c|}{0.0046} 
                    & \multicolumn{1}{c|}{0.6957} 
                    & \multicolumn{1}{c|}{0.67} 
                    & \multicolumn{1}{c|}{0.44} 
                    &  1
                    & \multicolumn{1}{c|}{0.0146} 
                    & \multicolumn{1}{c|}{0.2192} 
                    & \multicolumn{1}{c|}{0.73} 
                    & \multicolumn{1}{c|}{0.63} 
                    &  2 \\ 
                    \hline
                    \multicolumn{1}{|p{2cm}|}{randomAccess} 
                    & \multicolumn{1}{c|}{8.8987} 
                    & \multicolumn{1}{c|}{0.8896} 
                    & \multicolumn{1}{c|}{0.59} 
                    & \multicolumn{1}{c|}{0.49} 
                    &  1
                    & \multicolumn{1}{c|}{N/A} 
                    & \multicolumn{1}{c|}{N/A} 
                    & \multicolumn{1}{c|}{N/A} 
                    & \multicolumn{1}{c|}{N/A} 
                    &  N/A \\ 
                    \hline
\end{tabular}
\label{tab:cudatoopenmp}
\end{table*}

\section{Experiments and Results}


We experimented with LASSI on a Linux server equipped with two NVIDIA A100 GPUs, each with 40 GB of memory. The open-source models were hosted through a local deployment of Ollama \cite{ollama}, and GPT-4 was accessed through an API calling a private instance of the model. We experimented with several current open-source, code-centric LLMs available at the time of this work, and found that Codestral \cite{codestral}, Wizard Coder \cite{wizardcoder}, and DeepSeek Coder v2 \cite{deepseekcoder} performed sufficiently well to provide a viability demonstration of LASSI.

With the ten HeCBench applications, as outlined in Section \ref{sec:hecbench}, we sequentially ran the complete pipeline, covering 80 bi-directional translation scenarios between CUDA and OpenMP across ten applications and four LLMs. With each run, we captured compilation and execution results, code similarity metrics, and the runtime and standard output for those with successful execution.

\subsection{Evaluation Metrics}
\label{sec:metrics}

For the initial demonstration of LASSI's viability, we focus on basic metrics for the generated code, aiming to assess usability without delving into theoretical correctness or performance enhancements. Future work will include exploring additional metrics and refining prompt strategies. 

Tables \ref{tab:openmptocuda} and \ref{tab:cudatoopenmp} provide five metrics for each application code translation. 
 The first metric, \textit{Runtime}, provides the runtime of the LASSI-generated code.  The second metric, $Ratio$, is defined as the runtime of the original source code in the target programming language divided by the runtime of the LASSI-generated code. If \textit{N/A} is given, then the LASSI-generated code either failed to execute or its standard output did not match the expected result compared to the output of the source code.



Recognizing that code may be developed in more than one way to achieve the same solution, we do not expect the code generated by LASSI  to match the source code line-for-line. Nevertheless, evaluating the similarity between the source and LASSI-generated codes provides valuable insights for comparing performance across LLMs \cite{OpenMP_Fortran}. We include two string comparisons for \emph{code similarity}, corresponding to metrics three and four:
\begin{itemize}
\item $Sim\mbox{-}T$ is token-based, which  tokenizes both codes and uses a Ratcliff-Obershelp sequence comparison algorithm \cite{ratcliff} to find contiguous matching subsequences.  It generates a similarity ratio within [0, 1], with values over 0.6 indicating high similarity. 
\item $Sim\mbox{-}L$ is line-based, comparing codes line-by-line by counting identical lines regardless of order. The ratio represents the number of identical lines over the total lines in the longer code, with a higher ratio indicating more similarity, even if lines are in different order.
\end{itemize}

The final metric reported is $Self\mbox{-}corr$, corresponding to the number of self-correcting iterations the pipeline performed to re-prompt the LLM to correct compilation and execution errors. If the $Self\mbox{-}corr$ value is 0, then LASSI generated code that successfully compiled and executed on the first try. If this value is $> 0$ and the scenario also includes a $Ratio$, then the final generated code successfully compiled and executed, but the LLM required multiple self-corrections to obtain its code translation. 

\subsection{OpenMP to CUDA Translations}
We ran the automated pipeline configured to refactor codes developed in  OpenMP to CUDA. The selected HeCBench codes were input into the pipeline for translation. Also, the corresponding HeCBench codes in the target CUDA were compiled and executed with their standard outputs captured for visual inspection with the output of the translated code. The process for each code was repeated with four LLMs.

Table \ref{tab:openmptocuda} lists the results for our OpenMP to CUDA translations. If an LLM could not generate code that was compiled, executed,  or if the output was significantly different from the expected result, the record is marked as N/A. 

We observe that $80\%$ of the translations from OpenMP to CUDA successfully generated executable code with results similar to the source HeCBench code in CUDA.
This result strongly indicates the effectiveness of LASSI, which incorporates a novel prompting strategy, provided domain knowledge, and automated self-correction to translate OpenMP to CUDA. 
Of these successful generations, we observe 78.1\% execute with average runtimes within 10\% or faster than the average runtime of the source CUDA code. Taking the ratio of 0.6 as a heuristic measure for reasonable similarity between codes, the pipeline generated 40.6\% of the successful codes in the experimental set at this threshold or higher. Finally, the self-correction counts across all OpenMP to CUDA code translations remained quite low, with 65.6\% of the trials generating executable code on the first attempt.

\subsection{CUDA to OpenMP Translations}

Table \ref{tab:cudatoopenmp} lists the results for our CUDA to OpenMP translations. 
The results strongly validate the feasibility of LASSI. Specifically, 
 85\% of the translation samples  successfully generated executable code with similar output as the source. Among these, 61.8\% achieved average runtimes near or below those of the source OpenMP codes, 47.1\% generated heuristically similar codes, and 55.9\% generated executable code on the first attempt.

\subsection{Discussion}

We highlight two noteworthy findings from the experiments given in Tables \ref{tab:openmptocuda} and \ref{tab:cudatoopenmp} to shed light on the initial quality of the code generated by LASSI. First, we observe a likely lower-quality generated code in Codestral's translation of \emph{bsearch} from CUDA to OpenMP, which may necessitate additional self-correcting prompts. The code similarity measures are moderate, and the translation successfully executed on the first attempt without errors requiring correction. We compared standard outputs between the original HeCBench and translated codes, confirming identical results except for reported timings. However, the average runtime of the translated code over multiple runs is 20\(\times\) longer than that of the source code. Upon comparing the two codes, we noted the translated code only implements the default single thread, whereas the original source code explicitly sets 256 threads.

Second, we examine DeepSeek Coder's translation of \emph{atomicCost} from CUDA to OpenMP and observe over a 66\(\times\) speedup. Upon comparing the standard outputs of the HeCBench source and the translated code, we confirm identical results. 
The translated version appears to utilize several alternative approaches to parallelization, including thread limits, memory allocation, loop structures with fewer atomic operations, and timing methods.

These examples emphasize the known \emph{sensitivity} of existing LLMs \cite{DSPy} in generating content for which they are ill-trained and the \emph{opportunity} for enhancing these models through strategic prompting with domain knowledge and self-correction. Also, we anticipate the development of enhanced pipelines configured with prompted goals, such as improving performance or reducing energy consumption, as feasible extensions to our current architecture.

\section{Summary and Future Work}

In this work, we have prototyped an LLM-based automated self-correcting
pipeline, LASSI, for translating between parallel programming languages. The initial results of 
evaluating LASSI with different application codes across four LLMs demonstrate the effectiveness of LASSI for generating executable parallel codes, with 80\% of OpenMP to CUDA translations and 85\% of CUDA to OpenMP translations producing the expected output. We also observe approximately 78\% of OpenMP to CUDA translations and 62\% of CUDA to OpenMP translations execute within 10\% of or at a faster runtime than the original benchmark code in the same language.

We plan to explore several extensions to LASSI for generating verifiable and more performant codes. In particular, we will integrate code verification to automatically compare expected results, with feedback incorporated through another self-correcting cycle. 


\section*{Acknowledgment}
This work is supported in part by US National Science Foundation grant CCF-2413597. 
This material is based upon work supported by the U.S. Department of Energy, Office of
Science, under contract number DE-AC02-06CH11357, at Argonne National Laboratory. Thanks to Michael Papka for helping with naming the project. 

\end{document}